\documentclass[pra,notitlepage,twocolumn,amsmath,amssymb,aps,superscriptaddress]{revtex4-2}

\usepackage{amsmath,amssymb,amsfonts,bbm,times}
\usepackage[dvipsnames]{xcolor}
\usepackage[T1]{fontenc}
\usepackage{braket}
\usepackage[normalem]{ulem}
\usepackage[colorlinks=true,bookmarks=false,linkcolor=RoyalBlue,urlcolor=RoyalBlue,citecolor=RoyalBlue,breaklinks]{hyperref}
\usepackage{graphicx}
\usepackage{bm}
\usepackage{physics}

 \usepackage{soul}

\begin{document}

\title{Gaussian boson sampling with click-counting detectors}

\author{Gabriele Bressanini}
\affiliation{QOLS, Blackett Laboratory, Imperial College London, London SW7 2AZ, United Kingdom}
\author{Hyukjoon Kwon}
\affiliation{Korea Institute for Advanced Study, Seoul 02455, South Korea}
\author{M.S. Kim}
\affiliation{QOLS, Blackett Laboratory, Imperial College London, London SW7 2AZ, United Kingdom}
\affiliation{Korea Institute for Advanced Study, Seoul 02455, South Korea}

\begin{abstract}
Gaussian boson sampling constitutes a prime candidate for an experimental demonstration of quantum advantage within reach with current technological capabilities. The original proposal employs photon-number-resolving detectors, however the latter are not widely available. On the other hand, inexpensive threshold detectors can be combined into a single click-counting detector to achieve approximate photon number resolution.
We investigate the problem of sampling from a general multi-mode Gaussian state using click-counting detectors and show that the probability of obtaining a given outcome is related to a new matrix function which is dubbed as the Kensingtonian. We show how the latter relates to the Torontonian and the Hafnian, thus bridging the gap between known Gaussian boson sampling variants.
We then prove that, under standard complexity-theoretical conjectures, the model can not be simulated efficiently.
\end{abstract}
\maketitle

\section{Introduction}
Boson sampling, a computational problem introduced by Aaronson and Arkhipov \cite{AABS} and conjectured to be hard to simulate on a classical machine \cite{scheel_permanent}, constitutes a prime candidate for an experimental proof of quantum advantage using photons.
In its original formulation, the task consists of sampling from the output state  of a passive linear optical network (LON) fed with single-photon input states.
Several variants of the task that lie in the same complexity class have been proposed since. These are usually devised by considering different classes of input states, such as photon-added coherent states \cite{photon_added_coherent}, photon-added or photon-subtracted squeezed vacuum states \cite{photon_added_squeezed} and, more recently, non-Gaussian input states involving Kerr-type non-linearities \cite{nonlinearBS}. 
Most notably, Gaussian boson sampling (GBS) \cite{GBSOG} avoids the 
experimental hurdles of generating indistinguishable single-photon Fock states by using squeezed states of light as the non-classical resource needed to show quantum advantage \cite{GBScomplexity,Grier2022complexityof}. On top of being a more experimentally feasible alternative to prove quantum advantage on photonic platforms, GBS finds application in simulating molecular vibronic spectra \cite{vibronic}, predicting molecular docking configurations for drug design \cite{molecular_docking} and in graph-related problems such as finding dense subgraphs \cite{dense_subgraphs} and perfect matchings counting \cite{perfect_matchings_counting}.

Boson sampling variants may also be designed by considering different kinds of detection, such as Gaussian measurements \cite{BSGaussianMeasurement} and photo-counting detection schemes. Focusing on the latter, two classes of GBS experiments have been investigated thus far: the initial proposal of GBS \cite{detailedGBS} makes use of photon-number resolving (PNR) detectors, and only a couple of years later a GBS implementation utilizing threshold (on/off) detectors was suggested, as a less experimentally demanding alternative to show quantum advantage \cite{Torontonian}.
We remind the reader that on/off detectors can only detect the presence or absence of quantum light, while PNR detectors are, in principle, able to perfectly distinguish between any Fock states.
Threshold detectors, such as superconducting nanowires or avalanche photo-diodes that can be operated at room temperature \cite{single_photon_detector_rewiew}, are widely available and inexpensive. 
Hence, to lessen  experimental challenges, most GBS experiments up to date employed threshold detection \cite{jwpan1,jwpan2} and only more recently, a 216-mode Gaussian boson sampler utilizing PNR detectors was used to claim quantum computational advantage \cite{xanadu_GBS}.
\\
\\
On the other hand, improved classical algorithms that exploit photon collisional events to reduce the simulation’s overhead of GBS experiments employing thresholds detectors \cite{bulmer2022boundary} as well as the proposal of new classical spoofing strategies \cite{bulmer2022boundary,villalonga2021efficient,oh2022spoofing} motivate the development of larger scale experiments of increasing computational complexity.
To this end, the introduction of PNR detectors into the GBS scenario grants access to detection events with a much larger total photon number (thus increasing the sample space size exponentially), a regime which remains unachievable using on/off detectors.
As true PNR detectors are not always accessible, multiple on/off detectors are routinely combined in a multiplexed fashion to achieve approximate photon number resolution. 
The underlying idea of click-counting detection \cite{clickdetection} consists in dividing incoming light into weaker signals that are then measured  with threshold detectors, the final measurement output simply being the number of on/off detectors that registered the presence of photons. 
Hence, click-counting detectors can be seen as an intermediate case between threshold and PNR detectors.
We also recall that photon-number resolution is often  needed in GBS applications, e.g. to study higher-energy molecular vibronic transitions \cite{vibronic,vibronic_experimental}. 

In this work, we study the problem of sampling from a generic multi-mode Gaussian state using click-counting detectors, bridging the gap between GBS experiments utilizing PNR and threshold detectors, which can be seen as special instances of click-counting GBS. 
Besides fundamental interest, Gaussian boson samplers employing click-counting detectors have a clear experimental appeal, as it provides an easier way to achieve approximate photon number resolution.
In particular, we provide a closed-form expression for the probability of observing a given click-pattern outcome and show that it is related to a new matrix function which is dubbed as the Kensingtonian.
The latter plays an analogous role to the Hafnian and Torontonian in GBS variants employing PNR and on/off detection, respectively.
We show that, when the probability of observing two or more photons in each output mode is negligible, our model can not be efficiently simulated using a classical machine under standard complexity-theoretic conjectures, thus making the setup suitable to prove quantum advantage. In Table \ref{table} we present the matrix functions needed to compute the output probability distribution obtained when sampling from a Gaussian state using photo-counting measurements.

This paper is structured as follows. In Sec. \ref{sec_gaussian_states} we revise Gaussian states and their representations through phase-space quasi-probability distributions. In Sec. \ref{sec_old_GBS} we review the theoretical aspects of GBS experiments employing PNR and threshold detectors. 
In Sec. \ref{sec_click_detection} the click-counting detection scheme is introduced. Sections \ref{sec_ken} and \ref{sec_complexity_proof} are dedicated to the main findings of this work: we investigate the problem of sampling from a Gaussian state using click-counting detectors, we give the definition of the Kensigtonian and we prove the computational complexity of the sampling task.
Lastly, in Sec. \ref{sec_conclusions} we draw conclusions and give some final remarks.

\begin{table}[]
    \centering
    \begin{ruledtabular}
    \begin{tabular}{c|cc} 
        Detection & zero-mean Gaussian state & displaced Gaussian state \\
        \hline
        on/off & Torontonian  & loop Torontonian \\
        click &  Kensingtonian* & loop Kensingtonian* \\
        PNR &  Hafnian & loop Hafnian
    \end{tabular}
    \end{ruledtabular}
    \caption{Matrix functions used to compute the output probability distribution obtained from measuring multi-mode Gaussian states with on/off, click-counting and PNR detection.
    The symbol * denotes the functions that are first introduced in this paper.}
    \label{table}
\end{table}

\section{Gaussian States}
\label{sec_gaussian_states}
In this section we briefly review the key aspects of Gaussian states and their representation in terms of phase-space quasi-probability distributions (PQDs) \cite{serafini}.
Let us consider a continuous variables system made up of $M$-bosonic modes described by annihilation operators $a_j$ that satisfy the standard commutation relations $[a_j,a^\dagger_k]=\delta_{jk}$. We can then introduce quadrature operators for each mode, defined as $q_j = ({a_j+a_j^\dagger})/{\sqrt{2}}$ and $p_j = ({a_j-a_j^\dagger})/{i\sqrt{2}}$ (where we have set $\hbar = 1$), and arrange them into the following vector 
\begin{equation}
    \bm{r}=(q_1,p_1,\dots,q_M,p_M)^\intercal \, . 
    \label{ordering}
\end{equation}
A Gaussian state $\rho$ is completely characterized by its vector of first moments (displacement) $\bm{\overline{r}}$ and its covariance matrix $\sigma$, defined respectively as
\begin{equation}
\overline{\bm{r}}=\Tr{\rho \bm{r}} \, , 
\end{equation}
\begin{equation}
\sigma=\Tr{\rho \lbrace \bm{r}-\bm{\overline{r}}, (\bm{r}-\bm{\overline{r}})^\intercal \rbrace } \, .
\end{equation}
One can show that the $\bm{s}-$ordered PQD for a generic $M$-mode Gaussian state with covariance matrix $\sigma$ and vector of first moments $\bm{\alpha}$ is given by \cite{nonlinearBS}
\begin{equation}
    W_{\rho}^{(\bm{s})}(\bm{\beta})=\frac{2^M}{\pi^M\sqrt{\det{\bm{\sigma}-\bm{\tilde{s}}}}}e^{-2(\bm{\beta}-\bm{\alpha})^\intercal(\bm{\sigma}-\bm{\tilde{s}})^{-1}(\bm{\beta}-\bm{\alpha})} \, .
    \label{sPQD_Gaussian}
\end{equation}
Here $\bm{s}=(s_1,\dots,s_M)^\intercal$ is the vector of operator orderings, where $s_j \in\mathbb{R}$, and the matrix $\tilde{\bm{s}}$ is defined as 
\begin{equation}
    \bm{\tilde{s}}=\bigoplus_{j=1}^M s_j \mathbb{I}_2 \, .
\end{equation}
The conventions used in Eq.~\eqref{sPQD_Gaussian} are such that for a single-mode coherent state $\ket{\gamma}$ the covariance matrix is the identity matrix $\sigma=\mathbb{I}_2$ and the vector of first moments reads $\bm{\alpha}=(\Re{\gamma},\Im{\gamma})$.
Note that Eq.~\eqref{sPQD_Gaussian} is well-defined \textit{iff} $\bm{\sigma}-\bm{\tilde{s}} \geq 0$, otherwise the $\bm{s}-$PQD becomes more singular than a delta function.
The well-known Husimi Q-function, Wigner function and Glauber-Sudarshan P-function are retrieved respectively for $\bm{s}=-\mathbb{I}_M$, $\bm{s}=0$ and $\bm{s}=\mathbb{I}_M$.

\section{GBS with PNR and threshold detectors}
\label{sec_old_GBS}

In this section we briefly review the two most widely studied and experimentally relevant instances of GBS, i.e. the original proposal, which consists of sampling from a Gaussian state using PNR detectors, and its variant that employs more readily available threshold detectors.
Both implementations are thought to have the potential to show quantum advantage (at least in some regimes),  as sampling from their output probability distribution using classical algorithms can be shown to be a computationally hard task.

We remind the reader that a quantum measurement can be described in terms of a positive operator-valued measure (POVM), whose elements $\lbrace \Pi_x \rbrace$ satisfy the conditions $\Pi_x\geq0$ and $\sum_x\Pi_x = \mathcal{I}$ , where $\mathcal{I}$ is the identity operator on the Hilbert space. The probability $p(x)$ of obtaining a given outcome $x$ when measuring a state $\rho$ is given by the Born rule, i.e. $p(x)=\Tr{\rho \Pi_x}$.

The POVM elements of PNR detection are the projectors on Fock states, namely $\ketbra{k}$ with $k$ non-negative integer. This kind of detector provides perfect photon number resolution and can thus distinguish between any Fock states with no uncertainty.
It was shown \cite{GBSOG} that the photon-number statistics obtained from an $M-$mode Gaussian state with a null vector of first moments reads 
\begin{equation}
    p(\bm{k}) = \frac{\text{Haf}[XO_{(S)}]}{\sqrt{\det{\Sigma}} \, k_1!\dots k_M!} \, ,
    \label{GBS_PNR_prob}
\end{equation}
where $\bm{k}=(k_1,\dots,k_M)^\intercal$ are the detected photon numbers $k_i\geq0$, $\Sigma$ is the covariance matrix of the state's Q-function, $O_{(S)}$ is a matrix constructed from $O =1-\Sigma^{-1}$ according to the detection outcome (see Ref.~\cite{detailedGBS} for more details) and 
\begin{equation}
    X=\begin{bmatrix} 0 & \mathbb{I} \\ \mathbb{I} & 0 \end{bmatrix} \, .
\end{equation}
The matrix function appearing in Eq.~\eqref{GBS_PNR_prob} is called the Hafnian and is defined as follows
\begin{equation}
    \text{Haf}[A]= \sum_{\mu\in \text{PMP}}\prod_{j=1}^n A_{\mu(2j-1),\mu(2j)} \,\, ,
    \label{hafnian_definition}
\end{equation}
where $A$ is a $2n\times 2n$ complex matrix, and PMP is the set of perfect matching permutations.
Eq.~\eqref{GBS_PNR_prob}  can be generalized to the case of an $M-$mode Gaussian state with non-zero displacement by introducing the loop Hafnian matrix function \cite{detailedGBS}.

On the other hand, threshold detectors can only distinguish between the absence or presence of light, without being able to resolve the photon content of the state. In this case $k_i\in\lbrace 0,1 \rbrace$ and the POVM elements are $\Pi_0 = \ketbra{0}$ and $\Pi_1 = \mathcal{I}-\Pi_0$. 
It was shown (see Ref.~\cite{Torontonian} for details) that the probability distribution obtained from measuring a $M-$mode Gaussian state with threshold detectors reads
\begin{equation}
    p(\bm{k}) = \frac{\text{Tor}[O_{(S)}]}{\sqrt{\det{\Sigma}}} \, ,
    \label{GBS_threshold_probability}
\end{equation}
where
\begin{equation}
    \text{Tor}[A]= \sum_{Z\in P([n])} (-1)^{\vert Z\vert} \frac{1}{\sqrt{\det{\mathbb{I}-A_{(Z)}}}}
\end{equation}
is the Torontonian of a $2n\times 2n$ matrix $A$, $P([n])$ is the power set of $[n]=\lbrace 1,\dots,n\rbrace$ and $A_{(Z)}$ denotes a matrix constructed from $A$ by eliminating rows and columns according to the set $Z$.
Once again, it is possible to lift the zero-displacement constraint and generalize Eq.~\eqref{GBS_threshold_probability} by introducing the loop Torontonian matrix function \cite{looptorpaper}.

A relation between the Torontonian and the Hafnian can be established by noticing that the probability of obtaining a specific output pattern in a GBS experiment employing threshold detectors can also be computed by summing the probabilities of all detection events of a GBS experiment using PNR detectors that are compatible with that specific click pattern.
In particular, we say a PNR detection pattern $\bm{k}^\prime$ is compatible with a threshold detection pattern $\bm{k}$ if $k_i = 0 \implies k^\prime_i = 0$ and $k_i = 1 \implies k^\prime_i \geq 1$.
This observation leads to the following identity \cite{Torontonian}
\begin{equation}
    \text{Haf}[XO] = \frac{1}{n!} \frac{d^n}{d\eta^n} \text{Tor}[\eta O] \Big{\vert}_{\eta=0} \, ,
\end{equation}
where $O$ is a $2n\times 2n$ matrix.

\section{Click counting detection}
\label{sec_click_detection}
In this section we describe the click-detection scheme introduced in Ref.~\cite{clickdetection}, where a multiplexing setup and on/off detectors are employed to achieve approximate photon-number resolution. The main idea consists in splitting the incoming state into weaker signals (distributing the intensity uniformly among the output ports of the interferometer) that will then be measured using threshold detectors. 
\begin{figure}
    \centering
    \includegraphics[width=0.48\textwidth]{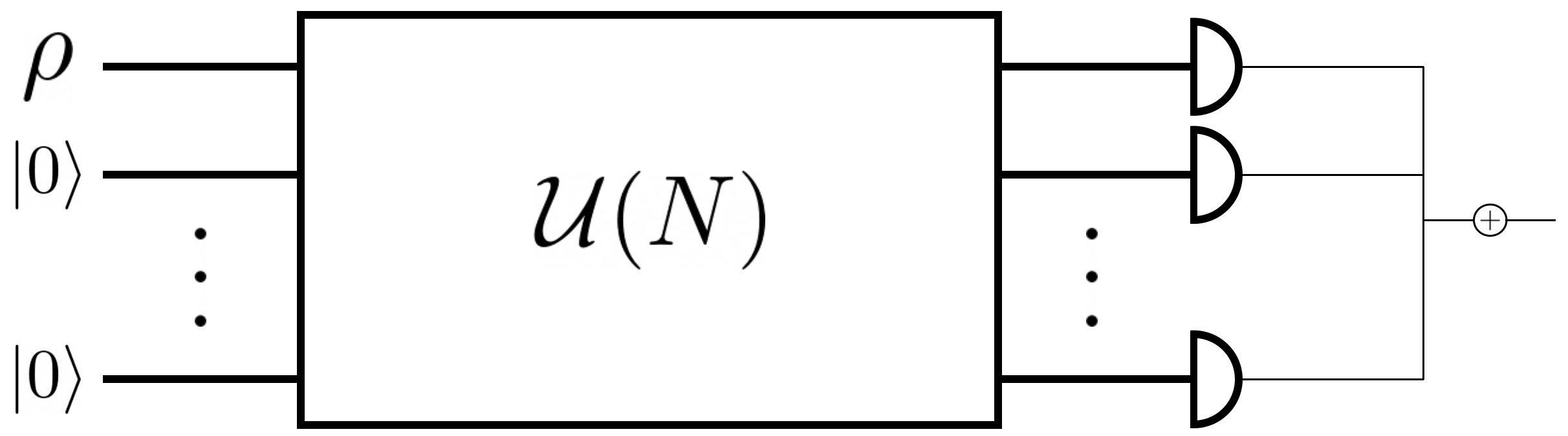}
    \caption{Schematics of a click-counting detector made up of $N$ threshold detectors. The quantum state $\rho$ to be measured enters an $N-$mode interferometer (represented by a unitary operation $\mathcal{U}(N)$) that splits the intensity equally among its output ports. $N$ on/off detectors then measure the output state and the detection results are summed into the final signal.}
    \label{fig_click_detection}
\end{figure}
Note how we are only interested in the total number of recorded clicks $k$, and not in the specific output pattern obtained in a given measurement.
In Fig.~\eqref{fig_click_detection} we display a schematic representation of a click-counting detector.
We remind the reader that the POVM elements of on/off detection are
\begin{equation}
    \Pi_0^{(1)} = \ketbra{0} = :e^{-\hat{n}}: \quad\quad \Pi_1^{(1)} = \mathcal{I} - \Pi_0^{(1)} \, ,
    \label{onoff_POVM_ideal}
\end{equation}
where $\mathcal{I}$ is the identity operator, $\hat{n}$ is the number operator and $:\bullet :$ denotes normal ordering of bosonic operators \cite{glauber}.
One can show that the POVM elements corresponding to a click-detector made up of $N$ on/off detectors are given by
\begin{equation}
    \Pi_{k}^{(N)} = :\binom{N}{k}e^{-\frac{N-k}{N}\hat{n}}(1-e^{-\frac{\hat{n}}{N}})^{k} : \, , 
    \label{click_POVM}
\end{equation}
where $k$, i.e. the number of recorded clicks, may vary between $0$ and $N$.
Note how we always refer to \textit{clicks} and not to photons.
As expected, for $N=1$ we retrieve the POVM elements of threshold detection, i.e.
\begin{equation}
    \Pi_k^{(1)} = : (e^{\hat{n}}-\mathcal{I})^k e^{-\hat{n}} : 
\end{equation}
with $k=0,1$. 
This formalism also allows us to easily introduce imperfections, such as sub-unit efficiency and finite dark count rate. To do so, we simply have to consider a click detector made up of noisy on/off detectors, whose POVM is obtained from Eq.~\eqref{onoff_POVM_ideal} by substitution of the number operator $\hat{n}$ with a suitable response function whose functional form depends on the specific noise model considered.
In this paper we consider the simple and widely used substitution $\hat{n}\mapsto\eta\hat{n}+\nu$, where $0\leq\eta\leq 1$ and $\nu\geq0$ are the efficiency and the dark count rate of the threshold detector, respectively, to obtain the following POVM elements
\begin{equation}
    \Pi_0^{(1)} = :e^{-(\eta\hat{n}+\nu)}: \quad\quad \Pi_1^{(1)}  = \mathcal{I} - \Pi_0^{(1)} \, .
    \label{noisy_onoff_POVM}
\end{equation}
One can then prove that the POVM elements of noisy click-counting detection are given by 
\begin{equation}
\Pi_k^{(N)}=\, : \binom{N}{k} \left[ e^{-\left(\eta \frac{\hat{n}}{N}+\nu \right)}\right]^{N-k}\left[ 1-e^{-\left(\eta \frac{\hat{n}}{N}+\nu \right)}\right]^k : \, .
\label{noisy_click_POVM}
\end{equation}
We note that this operators can be obtained from Eq.~\eqref{click_POVM} upon substituting $\hat{n}\mapsto\eta\hat{n}+N\nu$.
This is intuitively clear: we expect the click-counting detector to ``inherit'' the inefficiency of the threshold detectors, however the dark count rates from each on/off detector will add up to the ``total'' dark count rate of the click-counting detector.
The reason for this is that dark counts coming from different threshold detectors can be thought as independent Poisson variables and it is then well known that the sum of Poisson random variables is still a Poisson variable, whose mean value is the sum of the addends' mean values.
This also means that the performance of a noisy click-counting detector made up of $N$ threshold detectors characterized by Eq.~\eqref{noisy_onoff_POVM} should be compared to that of a PNR detector with sub-unit efficiency $\eta$ and dark count rate $N\nu$, whose related POVM elements read
\begin{equation}
    \Tilde{\Pi}_k = :\frac{(\eta\hat{n}+N\nu)^k}{k!} e^{-(\eta\hat{n}+N\nu)} :
    \label{noisy_PNR_POVM}
\end{equation}
In Appendix \ref{Appendix_bigNlimit} we show that $-$ as one might intuitively expect $-$ in the $N\rightarrow\infty$ limit and in the absence of noise we retrieve true PNR detection, i.e.
\begin{equation}
    \lim_{N\rightarrow\infty} \Pi_k^{(N)} =\, : \frac{\hat{n}^k}{k!}e^{-\hat{n}} : \,\equiv  \ketbra{k} \, .
    \label{convergence_click_to_PNR}
\end{equation}
The noisy case requires some additional attention, as the presence of non-zero dark count rate $\nu$ causes the expression to diverge if we are to take the formal limit.
In practice, typical values of a threshold detector's dark count rate are of the order $\nu\simeq 10^{-4}$, hence in regimes where simultaneously $N\gg1$ and $N\nu\ll1$ we still expect good convergence of Eq.~\eqref{noisy_click_POVM} to Eq.~\eqref{noisy_PNR_POVM}.

\section{The Kensingtonian}
\label{sec_ken}
In this section we derive the outcome probability distribution of an $M-$mode GBS experiment employing click-counting detectors, each made up of $N$ \textit{ideal} threshold detectors.
Note that for $N=1$ we are describing GBS with threshold detectors, and the probability distribution of outcomes is given by Eq.~\eqref{GBS_threshold_probability}. 
On the other hand, in the $N\rightarrow\infty$ limit we retrieve photon-number-resolving GBS with PNR detection, and the probability distribution converges to Eq.~\eqref{GBS_PNR_prob}. 
In this section we interpolate between these two special cases by deriving a closed formula valid for general $N$.

In click-counting GBS, a single detection event is denoted by $\bm{k}=(k_1,\dots,k_M)$, where $0\leq k_i\leq N$ $\forall i$. 
Note that the total number of clicks $n=\sum_i k_i$ is not fixed, as the number of photons entering the interferometer is not determined in the first place.
We want to compute 
\begin{equation}
    p(\bm{k})=\Tr{\rho \Pi_{\bm{k}}^{(N)}} \, ,
\end{equation}
where $\rho$ is a generic $M$-mode Gaussian state with covariance matrix $\sigma$ and null vector of first moments and $\Pi_{\bm{k}}^{(N)}$ is the POVM that characterize the $M$-mode click detection.
The latter simply reads
\begin{equation}
    \Pi_{\bm{k}}^{(N)} = \bigotimes_{i=1}^M \Pi_{{k_i}}^{(N)} \, ,
\end{equation}
where $\Pi_{{k_i}}^{(N)}$ is the POVM of a single click-counting detector, given by Eq.~\eqref{click_POVM}. The latter can also be expressed, by virtue of the binomial theorem, as 
\begin{equation}
     \Pi_{k_i}^{(N)} = \binom{N}{k_i}\sum_{\ell_i=0}^{k_i} \binom{k_i}{\ell_i}(-1)^{\ell_i} : e^{-\frac{N-k_i+\ell_i}{N}\hat{n}}: \, .
\end{equation}
Note that the operator $:e^{-\frac{N-k_i+\ell_i}{N}\hat{n}}:$ corresponds to the vacuum element $\Pi_0^{(1)}$ of the noisy on/off detection POVM Eq.~\eqref{noisy_onoff_POVM}, with an effective detection inefficiency given by $\lambda_i\equiv{(N-k_i+\ell_i)}/{N}\in [0,1]$.
This observation is crucial, as it implies that all is needed in order to obtain $p(\bm{k})$ are (noisy) vacuum statistics of marginal states of $\rho$.
Each of these contributions can be computed efficiently using the Gaussian formalism, hence $-$ as we will see $-$ the complexity of the sampling task arises from the exponential number of terms appearing in the expression of the probability.
We can express the latter as follows
\begin{equation}
    p(\bm{k}) = \pi^M \int d^{2M}\bm{\beta} \,  Q_{\rho}(\bm{\beta})\,  P_{\Pi_{\bm{k}}^{(N)}}(\bm{\beta}) \, ,
\end{equation}
where $Q_{\rho}(\bm{\beta})$ is the Husimi Q-function of $\rho$ and $P_{\Pi_{\bm{k}}^{(N)}}(\bm{\beta})$ is the P-function of the POVM element.
Using Eq.~\eqref{sPQD_Gaussian} with ordering parameter $s_j = -1\,\forall j$ we obtain 
\begin{equation}
    Q_\rho (\bm{\beta}) =  \frac{1}{\pi^M\sqrt{\det{{\Sigma}}}}e^{-\bm{\beta}^\intercal{\Sigma}^{-1}\bm{\beta}} \, ,
\end{equation}
where we have introduced the matrix ${\Sigma}=({{\sigma}+\mathbb{I}})/{2}$ for  convenience.
On the other hand, the P-function of $\Pi_{\bm{k}}^{(N)}$ is readily obtained once we know how to compute the P-function of $:e^{-\lambda_i\hat{n}}:$ .
One can prove that for $\lambda_i\neq 1$ 
\begin{equation}
    P_{:e^{-\lambda_i\hat{n}}:} (\beta_1,\beta_2) = \frac{1}{\pi(1-\lambda_i)}e^{-\frac{\lambda_i}{1-\lambda_i}(\beta_1^2+\beta_2^2)} \, .
\end{equation}
while for $\lambda_i = 1$ (i.e. for $\ell_i = k_i$) we have 
\begin{equation}
    P_{:e^{-\hat{n}}:}(\beta_1,\beta_2)=\delta(\beta_1)\delta(\beta_2) \, ,
\end{equation}
where $\beta_1$ and $\beta_2$ are the two (real) Cartesian coordinates of the complex plane.
After cumbersome calculations (see Appendix \ref{appendix_ken_derivation} for a detailed derivation) it is possible to show that the probability of observing a given click pattern $\bm{k}$ reads
\begin{equation}
    p(\bm{k}) = \frac{\text{Ken}[O]}{\sqrt{\det{\Sigma}}} \, .
    \label{clickGBSprob}
\end{equation}
Here $O = \mathbb{I}-\Sigma^{-1}$ and 
\begin{equation}
\begin{split}
    \text{Ken}[A] & = \sum_{0\leq \bm{d} \leq \bm{k}}
     \prod_{i=1}^M \left[ \binom{N}{N-k_i,k_i-d_i,d_i}(-1)^{k_i-d_i} \right] \\ & \prod_{j\notin Z}\left[\frac{N}{d_j} \right]
     \frac{1}{\sqrt{\det{(\mathbb{I}-A)_{(Z)}+D_Z}}} \, 
     \label{ken_def}
\end{split}
\end{equation}
is the \textit{Kensingtonian} of a $2M\times 2M$ matrix $A$.
$D_{Z}$ is a diagonal matrix defined as
\begin{equation}
    D_{Z}=\bigoplus_{i\notin Z} \left(\frac{N-d_i}{d_i} \right)\mathbb{I}_2 \, ,
    \label{Dmatrix}
\end{equation}
and the set $Z$ is defined as follows 
\begin{equation}
    Z=\lbrace i \vert 1\leq i \leq M ,\, d_i = 0 \rbrace \, .
\end{equation}
In Eq.~\eqref{ken_def} we have used the subscript notation $(\mathbb{I}-A)_{(Z)}$ to denote the matrix obtained from $(\mathbb{I}-A)$ by eliminating the corresponding rows/columns according to the set $Z$. In particular, if $Z=\lbrace a,b \rbrace$ we remove the rows(columns) numbered $2a-1,2a,2b-1$ and $2b$.
In Appendix \ref{appendix_loop_ken} we generalize Eq.~\eqref{clickGBSprob} to  displaced Gaussian input states by introducing the \textit{loop Kensingtonian} matrix function.

For $N=1$, Eq.~\eqref{clickGBSprob} must coincide with the probability distribution of a GBS experiment employing threshold detectors Eq.~\eqref{GBS_threshold_probability}.
In Appendix \ref{appendix_ker_to_tor} we explicitly prove the following identity that links the Kensingtonian and the Torontonian
\begin{equation}
    \text{Ken}[A] \stackrel{\small{N=1}}{=} \text{Tor}[A_{(\mathcal{K})}] \, .
\end{equation}
On the other hand, the outcome probability distribution of click-counting GBS Eq.~\eqref{clickGBSprob} converges to that of PNR GBS Eq.~\eqref{GBS_PNR_prob} in the $N\rightarrow \infty$ limit, as a consequence of the fact that in that same limit the click-counting detection POVM converges to that of true PNR detection. This in turn implies that the Hafnian can be retrieved as a limiting case of the Kensingtonian. We leave it as an open question whether this can be exploited to develop faster algorithms to approximate the Hafnian of a matrix.

One might also be interested in knowing the regimes where the photon counting statistics coming from a GBS experiment employing click-counting detectors well approximates Eq.~\eqref{GBS_PNR_prob}. 
Of course, the convergence of the click-counting detector's POVM to that of the PNR detector implies that we can always increase $N$ to make the total variational distance (TVD) between the two distributions arbitrarily small.
However, in what follows we want to give a more quantitative and practical indication of the number of threshold detectors needed in order to have a good agreement between Eq.~\eqref{clickGBSprob} and Eq.~\eqref{GBS_PNR_prob}.
This is particularly relevant for GBS experiments aimed at applications (rather than those aimed at proving quantum advantage), where the Hafnian's specific functional form and properties are crucial to map the sampling task onto problems in graph theory and chemistry.

\begin{figure}
    \centering
    \includegraphics[width=0.48\textwidth]{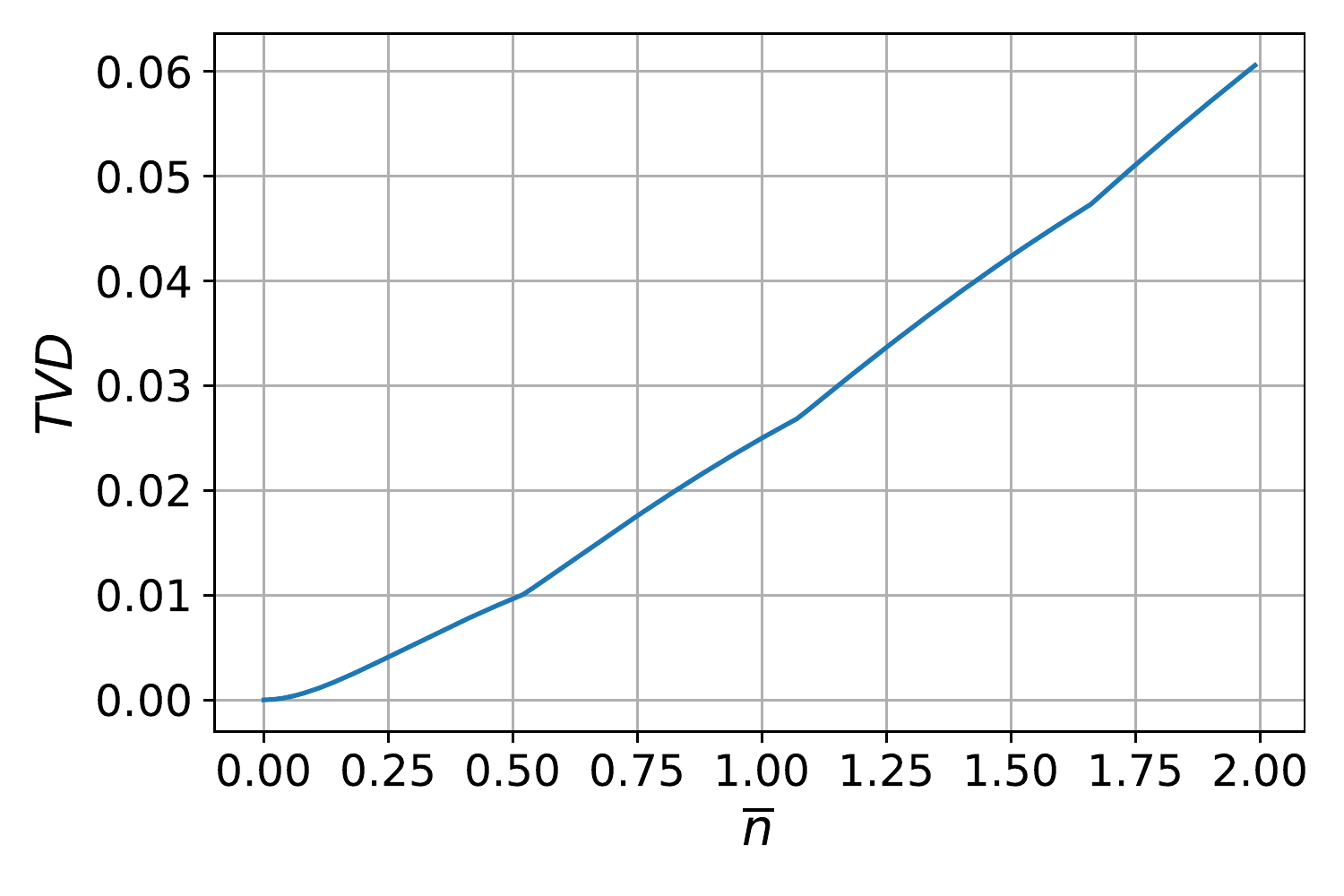}
    \caption{The total variational distance (TVD) between the probability distributions coming from a PNR and click-counting detector made up of $N=8$ threshold detectors, as a function of the mean photon number $\overline{n}$ of the probe thermal state $\nu_{th}(\overline{n})$.
    In the energy regime of interest we considered there is a good agreement, with TVD smaller than 0.05.}
    \label{fig_tvd}
\end{figure}

We remind the reader that, on average, a Haar-random LON equally distributes the intensity among its output modes, and denote with $\overline{n}$ the average photon density per mode.
Most GBS experiments up to date, especially those targeting  applications, have modest values of photon densities per output mode, usually $\overline{n}<1$.
Furthermore, tracing out all output modes but one leaves us with a Gaussian marginal state that approximately looks like a thermal state $\nu_{th}(\overline{n})$.
We thus restrict ourselves to this single-mode case and numerically compute the TVD to give a rough estimate of how many threshold detectors making up a click-detector are needed so that the latter well approximates an ideal PNR detector in the energy regime of interest.
The two probability distributions needed to compute this TVD are given by Eq.~\eqref{GBS_PNR_prob} and Eq.~\eqref{clickGBSprob}, with $M=1$ and $\sigma = (2\overline{n}+1)\mathbb{I}_2$.
In Fig.~\eqref{fig_tvd} we show how using just $N=8$ threshold detectors (corresponding to 3 multiplexing steps, if we are using the simple scheme where intensity is halved at each layer made up of balanced beamsplitters) already provides a good agreement between the two distributions.

\section{Complexity of click-GBS}
\label{sec_complexity_proof}
In this section we prove that sampling from the output probability distribution of a Gaussian boson sampler utilizing click-counting detectors constitutes a hard problem to solve on a classical computer.
Proofs of hardness of the Boson sampling task and its variants rely on the assumption that the probability of observing a collision (i.e. two or more photons) in any of the interferometer's output modes is negligible. 
This is also known as the non-collisional regime.
In Ref.~\cite{AABS} the authors proved that the probability $\varepsilon$ of observing at least a collision at the output of an $M-$mode Haar-random LON fed with $n$ single photons can be bounded as follows
\begin{equation}
    \expval{\varepsilon}_\mathcal{U} \leq \frac{2n^2}{M} \, ,
    \label{collision_probability_bound}
\end{equation}
where $\expval{\cdot}_\mathcal{U}$ denotes averaging over $M\times M$ Haar-random unitary matrices. In GBS the number of photons is not fixed, hence the squared number of photons $n^2$ on right-hand side of Eq.~\eqref{collision_probability_bound} needs to be further averaged  according to the photon number probability distribution of the Gaussian state.
This means that it is possible to set the collision probability to be a small constant by choosing the scaling of the number of modes appropriately.

Intuitively, in this regime, the statistics coming from PNR, threshold and click-counting detectors should all behave similarly. In the following we rigorously formalize this idea, by adjusting the arguments used in Ref.~\cite{Torontonian} to prove that sampling from a Gaussian boson sampler with on/off detectors is a computationally-hard problem. 
Roughly speaking, the authors proved that $-$ when collision probability is small $-$ an approximation of threshold GBS would also constitute a good approximation of GBS with PNR detectors, an event considered to be unlikely as it would imply the collapse of the polynomial hierarchy to the third level. Hence, it is concluded that $-$ under standard complexity-theoretic arguments $-$ threshold GBS is in the same complexity class as the original GBS proposal.

It thus suffices to show that in the non-collisional regime the output probability distribution $p$ of click-counting GBS is arbitrarily close to that of threshold GBS, which we denote with the symbol $\tilde{p}$.
We define the set of collision events of an $M-$mode sampling problem as
\begin{equation}
    \mathcal{C} = \lbrace \bm{k}=(k_1,\dots,k_M) \,\vert\, k_i>1 \text{ for some } i \in \lbrace1,\dots,M\rbrace\rbrace \, .
\end{equation}
By definition we have that $\tilde{p}(\bm{k}\in\mathcal{C})=0$, as threshold detectors can either click once or not click at all.
On the other hand, the probability of observing a collision event for click-counting GBS reads
\begin{equation}
    \varepsilon = \sum_{\bm{k}\in\mathcal{C}} p(\bm{k}) \, .
\end{equation}
The total variational distance between $p$ and $\tilde{p}$ reads
\begin{equation}
\begin{split}
    \vert\vert p - \tilde{p} \vert\vert_1  & = \frac{1}{2}\sum_{\bm{k}} \vert p(\bm{k})-\tilde{p}(\bm{k})\vert \\ & =  \frac{1}{2}\sum_{\bm{k}\in\mathcal{C}} \vert p(\bm{k})-\tilde{p}(\bm{k})\vert + \frac{1}{2}\sum_{\bm{k}\notin\mathcal{C}} \vert p(\bm{k})-\tilde{p}(\bm{k})\vert  \\ & = \frac{1}{2}\sum_{\bm{k}\in\mathcal{C}}  p(\bm{k})+ \frac{1}{2}\sum_{\bm{k}\notin\mathcal{C}} \vert p(\bm{k})-\tilde{p}(\bm{k})\vert \\ & = \frac{\varepsilon}{2} + \frac{1}{2}\sum_{\bm{k}\notin\mathcal{C}} \vert p(\bm{k})-\tilde{p}(\bm{k})\vert \, . 
\end{split}
\label{tvd}
\end{equation}
By close inspection of the click-counting detection POVM Eq.~\eqref{click_POVM} we notice that $\Pi^{(N)}_0 = :e^{-\hat{n}}: = \ketbra{0} \quad$ for every value of $N$.
Using this and the fact that the elements of a POVM resolve the identity we can write
\begin{equation}
     \sum_{k=1}^N \Pi_k^{(N)} = \mathcal{I} - \ketbra{0} = \Pi_1^{(1)} \, .
\end{equation}
This implies that the probability of a threshold detector clicking is equal to the probability of a click detector detecting any number of clicks between $1$ and $N$.
Generalizing this to $M$ modes we obtain 
\begin{equation}
    \tilde{p}(\bm{k}) = p(\bm{k}) + \sum_{\bm{k}^\prime\in\mathcal{C}_{\bm{k}}} p(\bm{k}^\prime) \, ,
    \label{niceformula}
\end{equation}
where $\bm{k}\notin\mathcal{C}$ is a collision-less detection event and $\mathcal{C}_{\bm{k}}$ is the set of all possible collision events compatible with $\bm{k}$
\begin{equation}
    \mathcal{C}_{\bm{k}}=\lbrace \bm{k}^\prime \in \mathcal{C} \,\vert\,\text{if } k_i = 0 \implies k_i^\prime = 0 \rbrace \, .
\end{equation}
If we now substitute Eq.~\eqref{niceformula} into Eq.~\eqref{tvd} we obtain
\begin{equation}
    \begin{split}
    \vert\vert p - \tilde{p} \vert\vert_1  & = \frac{\varepsilon}{2} + \frac{1}{2}\sum_{\bm{k}\notin\mathcal{C}} \vert p(\bm{k})-\tilde{p}(\bm{k})\vert  \\ & = \frac{\varepsilon}{2} + \frac{1}{2}\sum_{\bm{k}\notin\mathcal{C}} \sum_{\bm{k}^\prime\in\mathcal{C}_{\bm{k}}} p(\bm{k}^\prime) \\&= \frac{\varepsilon}{2}+\frac{1}{2} \sum_{\bm{k}^\prime\in\mathcal{C}}p(\bm{k}^\prime) \\&= \varepsilon\, . 
\end{split}
\end{equation}
Let us now consider a probability distribution $\pi$ that approximates $p$ arbitrarily well, i.e. it satisfies $\vert\vert p-\pi \vert\vert_1 = \varepsilon^\prime$. 
Using the triangle inequality of the $L^1$ norm we can bound $\vert\vert \tilde{p}-\pi \vert\vert_1$ as follows
\begin{equation}
    \vert\vert \tilde{p}-\pi \vert\vert_1 = \vert\vert \tilde{p} - p + p -\pi \vert\vert_1 \leq \vert\vert \tilde{p}-p\vert\vert_1 + \vert\vert p-\pi \vert\vert_1 = \varepsilon + \varepsilon^\prime \, .
    \label{maininequality}
\end{equation}
If we now assume that there exists a polynomial time algorithm that can sample from $\pi$, then Eq.~\eqref{maininequality} tells us that the same algorithm can sample efficiently from an arbitrarily good approximation of $\tilde{p}$, thus causing the collapse of the polynomial hierarchy to the third level and concluding our proof of hardness for click-counting GBS.
In Appendix \ref{Appendix_alternative_proof} we outline an alternative proof of hardness that does not make use of the output probability distribution of a Gaussian boson sampler employing threshold detectors.

We can now discuss the time complexity of computing the Kensingtonian according to its definition Eq.~\eqref{ken_def}. Let us first call $n=\sum_i k_i$ the total number of clicks. It is clear that, similarly to the Torontonian, the complexity of computing the Kensingtonian arises from the number of determinant contributions one needs to evaluate, which is given by 
\begin{equation}
    \sum_{0\leq\bm{d}\leq\bm{k}} = \prod_{i=1}^M (k_i + 1) \equiv F(\bm{k})
    \label{how_many_det} \, ,
\end{equation}
where $0\leq k_i\leq N$.
For $N=1$ (threshold detectors) we have $k_i = 0,1$, hence $\prod_{i=1}^M(k_i + 1) = 2^n$. In this special case the result depends solely on the total number of clicks $n$, however one can easily see that this is not the case for $N>1$, as we would need and increasing number of functions of all the $k_i$ to completely characterize the expression, making it impractical. 
An exact, simple, closed formula is thus out of reach, however in the following we show that, on average, we still obtain an exponential scaling in $n$.

The average value of $F(\bm{k})$ at fixed $n$ can be expressed as a sum of a collision-less term and collision term, namely
\begin{equation}
    \langle F \rangle_n = \sum_{\bm{k}\notin\mathcal{C}_n}p(\bm{k})F({\bm{k}}) + \sum_{\bm{k}\in\mathcal{C}_n}p(\bm{k})F({\bm{k}}) \, .
    \label{mean_F}
\end{equation}
Here $\mathcal{C}_n$ is the set of collision detection events at a fixed total number of clicks and we use the symbol $\expval{\cdot}_n$ to denote averaging at constant $n$.
When there are no collisions we simply have that $F(\bm{k})=2^n$.  We can thus write
\begin{equation}
     \langle F \rangle_n =(1-\tilde{\varepsilon})2^n + \sum_{\bm{k}\in\mathcal{C}_n}p(\bm{k})F({\bm{k}}) \, ,
\end{equation}
where $\tilde{\varepsilon}$ is the probability of observing a collision at fixed $n$, namely 
\begin{equation}
    \tilde{\varepsilon} =\sum_{\bm{k}\in\mathcal{C}_n}p(\bm{k}) \, .
\end{equation}
It can easily be seen that   $F(\bm{k})\geq(n+1)$, hence
\begin{equation}
     \langle F \rangle_n \geq (1-\tilde{\varepsilon})2^n + \tilde{\varepsilon} (n+1) \, ,
\end{equation}
which confirms the exponential scaling of the number of terms to be evaluated in Eq.~\eqref{ken_def} with the total number of clicks.
Analogously, we can upper bound $F(\bm{k})$ using the inequality of the arithmetic and geometric means and show that the following chain of inequalities holds
\begin{equation}
    (1-\tilde{\varepsilon})2^n + \tilde{\varepsilon} (n+1) \, \leq \langle F \rangle_n \leq (1-\tilde{\varepsilon})2^n + \tilde{\varepsilon} e^n \, .
\end{equation}
We recall that the determinants present in the definition of the Kensingtonian can be computed efficiently using the standard algorithm based on the Cholesky decomposition, whose time complexity scales with the cube of the matrix dimension. 
Hence it follows that a direct evaluation of Eq.~\eqref{ken_def}, in the non-collisional regime, leads to a complexity upper bounded by $O(n^3 2^n)$. 
We emphasize that we do not claim optimality, and we leave it as an open question the possibility of exploiting the structure of the Kensingtonian to find faster algorithms for its evaluation.

\section{Conclusions}
\label{sec_conclusions}
In this paper we have investigated the problem of sampling from Gaussian states with click-counting detectors and found a closed-form expression for the probability distribution.
The latter is related to the Kensingtonian, a new matrix function that plays an analogous role to the Hafnian and the Torontonian in GBS experiments employing PNR and threshold detectors, respectively. We then proved that, in the non-collisional regime, the problem at study still gives rise to a computationally hard problem, intractable using classical sampling algorithms, and showed how the Kensingtonian is related to known matrix functions in limiting cases of interest.

Our work leaves some open questions. 
We recall that Eq.~\eqref{clickGBSprob} converges to Eq.~\eqref{GBS_PNR_prob} in the $N\gg1$ limit, hence it would be interesting to investigate whether the Kensingtonian's structure could be exploited to design new algorithms to approximate the Hafnian.
Future efforts will also focus on studying the classical simulability of a GBS experiment employing noisy click-counting detection, where we envision the existence of a trade-off relation between $N$, i.e. the number of threshold detectors making up a single click-counting detector, and the noise parameters that characterize each on/off detector, for the system to enter a regime where achieving quantum advantage is not ruled out.
\\
\\
\textit{Note added} $-$ As we are finalizing the manuscript we became aware of a recent experiment \cite{deng2023gaussian} reporting an implementation of a Gaussian boson sampling device employing \textit{unbalanced} click-counting detectors,
i.e. the intensity of the incoming light is split unevenly among the threshold detectors that make up a single click-counting detector.
Our theoretical modeling and that presented in Ref.~\cite{deng2023gaussian} are consistent with each other, as they originate from the same POVM describing a click-counting detector.
In this paper, we have exploited the structure of the POVM to obtain a closed analytical formula for the outcome probability distribution, that can be readily applied to an $M-$mode GBS task employing balanced click-counting detectors.
In particular, we showed that the probability of a particular detection outcome may be obtained by computing the Kensingtonian of a $2M\times 2M$ matrix.
On the other hand, the authors of Ref.~\cite{deng2023gaussian} model their experimental setup as larger instance of a $MN-$mode GBS experiment employing on/off detectors.
As a result, they sum over the probabilities of (combinatorially-many) threshold-detection outcomes that give rise to the same click-detection pattern.
Each of these probabilities requires the evaluation of the Torontonian of matrices that are up to $2MN \times 2MN$ in dimension, and may therefore lead to a less efficient computation of a click-detection outcome probability with respect to the approach presented in this paper.

\section{Acknowledgments}
G.B. is part of the AppQInfo MSCA ITN which received funding from the European Union’s Horizon 2020 research and innovation programme under the Marie Sklodowska-Curie grant agreement No 956071.
G.B. thanks Jan Sperling for introducing him to click-counting detection and Francesco Vigan\`o for helpful discussions.
The authors thank Changhun Oh for providing useful comments.
H. K. is supported by the KIAS Individual Grant No. CG085301 at Korea Institute for Advanced Study.
MSK acknowledges the KIST Open Research Programme, Samsung GRC programme and the KIAS visiting professorship.

\bibliography{biblio}

\onecolumngrid

\appendix

\section{Click-counting probability}
\label{appendix_ken_derivation}
In this section we derive the probability of observing a given click-counting pattern $\bm{k}=(k_1,\dots,k_M)$ when a generic $M-$mode Gaussian state $\rho$ with covariance matrix $\sigma$ and null vector of first moments is sampled using $M$ click-counting detectors, each composed of $N$ threshold detectors. We remind the reader that each click-detector can measure at most $N$ clicks.
The probability reads
\begin{equation}
    p(\bm{k})=\Tr{\rho \Pi_{\bm{k}}^{(N)}} \, ,
\end{equation}
where $\Pi_{\bm{k}}^{(N)}$ is the POVM element that characterize the $M$-mode click detection
\begin{equation}
    \Pi_{\bm{k}}^{(N)} = \bigotimes_{i=1}^M \Pi_{{k_i}}^{(N)} \, .
\end{equation}
Here $\Pi_{{k_i}}^{(N)}$ is the POVM of a single click-counting detector and is given by
\begin{equation}
    \Pi_{k_i}^{(N)} = :\binom{N}{k_i}e^{-\frac{N-k_i}{N}\hat{n}}(1-e^{-\frac{\hat{n}}{N}})^{k_i} : =\binom{N}{k_i}\sum_{\ell_i=0}^{k_i} \binom{k_i}{\ell_i}(-1)^{\ell_i} : e^{-\frac{N-k_i+\ell_i}{N}\hat{n}}: \, .
\end{equation}
Note that the operator $: e^{-\frac{N-k_i+\ell_i}{N}\hat{n}}:$ corresponds to the vacuum element $\Pi_0^{(1)}$ of the POVM associated with noisy threshold detection with effective detection inefficiency given by $\lambda_i\equiv({N-k_i+\ell_i})/{N}\in [0,1]$.
In particular, $:e^{-\lambda_i\hat{n}}:$ represents an unnormalized thermal state for $\lambda_i\in(0,1)$, the vacuum state for $\lambda_i=1$ and the identity operator for $\lambda_i = 0$, respectively.
The probability can be expressed as 
\begin{equation}
    p(\bm{k}) = \pi^M \int d^{2M}\bm{\beta} \,  Q_{\rho}(\bm{\beta})\,  P_{\Pi_{\bm{k}}^{(N)}}(\bm{\beta}) \, ,
    \label{prob_Q_and_P_func}
\end{equation}
where $Q_{\rho}(\bm{\beta})$ is the Husimi Q-function of $\rho$ and $P_{\Pi_{\bm{k}}^{(N)}}(\bm{\beta})$ is the P-function of $\Pi_{\bm{k}}^{(N)}$.
Using Eq.~\eqref{sPQD_Gaussian} with ordering parameters $s_j = -1\,\forall j$ we obtain
\begin{equation}
    Q_\rho (\bm{\beta}) =  \frac{1}{\pi^M\sqrt{\det{{\Sigma}}}}e^{-\bm{\beta}^\intercal{\Sigma}^{-1}\bm{\beta}} \, ,
\end{equation}
where we have introduced the matrix ${\Sigma}=({{\sigma}+\mathbb{I}})/{2}$.
\\
In order to compute the P-function of $\Pi_{\bm{k}}^{(N)}$ we just need the P-function of the unnormalized thermal state $:e^{-\lambda_i\hat{n}}:$ .
Using Eq.~\eqref{sPQD_Gaussian} one can prove that for $\lambda_i\neq 1$
\begin{equation}
    P_{:e^{-\lambda_i\hat{n}}:} (\beta_1,\beta_2)= \frac{1}{\pi(1-\lambda_i)}e^{-\frac{\lambda_i}{1-\lambda_i}(\beta_1^2+\beta_2^2)} =  
    \frac{N}{\pi(k_i-\ell_i)}e^{-\frac{N-k_i+\ell_i}{k_i-\ell_i}(\beta_1^2+\beta_2^2)}\, .
\end{equation}
For $\lambda_i = 1$ (i.e. for $\ell_i = k_i$) we have $ P_{:e^{-\hat{n}}:}(\beta_1,\beta_2)=\delta(\beta_1)\delta(\beta_2)$.
In particular, in the following we will always understand the term
\begin{equation}
        \frac{N}{\pi(k_i-\ell_i)}e^{-\frac{N-k_i+\ell_i}{k_i-\ell_i}(\beta_1^2+\beta_2^2)}
\end{equation}
to be the delta function $\delta(\beta_1)\delta(\beta_2)$ whenever $\ell_i = k_i$. In fact, recall that 
\begin{equation}
    \lim_{\varepsilon\rightarrow 0} \frac{1}{\pi \varepsilon^2}e^{-(x^2+y^2)/{\varepsilon^2}} = \delta(x)\delta(y) \, .
\end{equation}
Note that throughout this work we will use Cartesian coordinates of the complex plane, meaning that $\beta_1$ and $\beta_2$ are real variables.
Now recall that the P-function of a tensor product is simply the product of the P-functions, i.e.
\begin{equation}
    P_{\Pi_{\bm{k}}}^{(N)}(\bm{\beta}) = \prod_{i=1}^M P_{\Pi_{k_i}}^{(N)}(\beta_{2i-1},\beta_{2i}) \, .
\end{equation}
After some calculations, one obtains
\begin{equation}
    P_{\Pi_{k_i}^{(N)}}(\beta_{2i-1},\beta_{2i}) = \binom{N}{k_i}\sum_{\ell_i = 0}^{k_i} \binom{k_i}{\ell_i}(-1)^{\ell_i} \frac{N}{\pi(k_i - \ell_i)} e^{-\frac{N-k_i+\ell_i}{k_i - \ell_i}(\beta_{2i-1}^2+\beta_{2i}^2)} \, , 
\end{equation}
i.e. a linear combination of Gaussians.
Note that, in the previous expression, for $\ell_i = k_i$ we obtain a delta function, which we will need to take care of separately.
Consequently, the probability reads 
\begin{equation}
    p(\bm{k}) = \int d^{2M}\bm{\beta} \, \frac{ e^{-\bm{\beta}^\intercal{\Sigma}^{-1}\bm{\beta}}  }{\sqrt{\det{{\Sigma}}}} \prod_{i=1}^M \binom{N}{k_i}\sum_{\ell_i = 0}^{k_i} \binom{k_i}{\ell_i} \frac{N(-1)^{\ell_i}}{\pi(k_i - \ell_i)} e^{-\frac{N-k_i+\ell_i}{k_i - \ell_i}(\beta_{2i-1}^2+\beta_{2i}^2)} \, .
\end{equation}
First, we want to invert the product and the sum that appear in the previous expression. This is easily done by using the following property
\begin{equation}
    \prod_{i=1}^M \sum_{\ell_i = 0}^{k_i} f(\ell_i,k_i) = \sum_{0\leq \ell_1 \leq k_1} \cdots \sum_{0\leq \ell_M \leq k_M}  \prod_{i=1}^M f(\ell_i,k_i) \equiv \sum_{0\leq \bm{\ell} \leq \bm{k}}  \prod_{i=1}^M f(\ell_i,k_i) \, , 
\end{equation}
where $f(\ell_i,k_i)$ is a generic function of $\ell_i$ and $k_i$. 
The probability then reads
\begin{equation}
\begin{split}
    p(\bm{k})  & = \sum_{0\leq \bm{\ell} \leq \bm{k}} \int d^{2M}\bm{\beta} \, \frac{ e^{-\bm{\beta}^\intercal{\Sigma}^{-1}\bm{\beta}}  }{\sqrt{\det{{\Sigma}}}} \prod_{i=1}^M \left[ \binom{N}{k_i} \binom{k_i}{\ell_i} \frac{N(-1)^{\ell_i}}{\pi(k_i - \ell_i)} e^{-\frac{N-k_i+\ell_i}{k_i - \ell_i}(\beta_{2i-1}^2+\beta_{2i}^2)} \right] 
    \\ & =  \sum_{0\leq \bm{\ell} \leq \bm{k}} \prod_{i=1}^M \left[ \binom{N}{k_i} \binom{k_i}{\ell_i} \frac{N(-1)^{\ell_i}}{\pi(k_i - \ell_i)}\right] \int d^{2M}\bm{\beta} \, \frac{ e^{-\bm{\beta}^\intercal{\Sigma}^{-1}\bm{\beta}}  }{\sqrt{\det{{\Sigma}}}} 
    e^{{\sum_{i=1}^M-\frac{N-k_i+\ell_i}{k_i - \ell_i}(\beta_{2i-1}^2+\beta_{2i}^2)}} \, .
 \end{split}
\end{equation}
To further ease the notation we introduce the multinomial coefficient
\begin{equation}
    \binom{N}{\ell_i}\binom{\ell_i}{k_i}
    = \frac{N!}{(N-\ell_i)!(\ell_i - k_i)!k_i!}
    \equiv \binom{N}{N-\ell_i,\ell_i-k_i,k_i}
\end{equation}
and make the variable change $d_i = k_i - \ell_i$.
\begin{equation}
\begin{split}
    p(\bm{k})  =  \sum_{0\leq \bm{d} \leq \bm{k}} \prod_{i=1}^M \left[ \binom{N}{N-k_i,k_i-d_i,d_i} \frac{N(-1)^{k_i-d_i}}{\pi d_i}\right] \int d^{2M}\beta \, \frac{ e^{-\bm{\beta}^\intercal{\Sigma}^{-1}\bm{\beta}}  }{\sqrt{\det{{\Sigma}}}} e^{ {\sum_{i=1}^M-\frac{N-d_i}{d_i}(\beta_{2i-1}^2+\beta_{2i}^2)}}
 \end{split}
\end{equation}
Let us also define the set
\begin{equation}
    Z=\lbrace i \vert 1\leq i \leq M ,\, d_i = 0 \rbrace \, ,
    \label{Z_def}
\end{equation}
which identifies the presence of delta functions in the integrand.
Integrating over $\delta(\beta_{2i-1})\delta(\beta_{2i})$ has the effect of setting $\beta_{2i-1}=\beta_{2i}=0$ which, in turn, is equivalent to deleting the corresponding rows and columns from the matrix $\Sigma^{-1}$. We will denote this new matrix with $(\Sigma^{-1})_{(i)}$. In general, we will use the notation $(\Sigma^{-1})_{(Z)}$ to denote the matrix obtained from $\Sigma^{-1}$ by eliminating the corresponding rows/columns according to the set $Z$. In particular, if $Z=\lbrace a,b \rbrace$ we will eliminate the rows(columns) numbered $2a-1,2a,2b-1$ and $2b$.
Note that, if we delete all the $2M$ rows and columns from $\Sigma^{-1}$ we are left with 1 by definition. After integrating over all the delta functions we are left with
\begin{equation}
    p(\bm{k})  =  \sum_{0\leq \bm{d} \leq \bm{k}}  \prod_{i=1}^M \left[ \binom{N}{N-k_i,k_i-d_i,d_i}(-1)^{k_i-d_i} \right]\prod_{i\notin Z}\left[\frac{N}{\pi d_i} \right]\int d^{2(M-\vert Z\vert)}\bm{\beta} \frac{1}{\sqrt{\det{\Sigma }}}e^{-\bm{\beta}^\intercal[(\Sigma^{-1})_{(Z)}+D_{Z}]\bm{\beta}} \, ,
 \label{newprob}
\end{equation}
where $\vert Z\vert$ is the cardinality of $Z$ and $D_Z$ is a diagonal matrix defined as
\begin{equation}
    D_Z=\bigoplus_{i\notin Z} \left(\frac{N-d_i}{d_i} \right)\mathbb{I}_2 \, .
\end{equation}
The remaining integrals in Eq.~\eqref{newprob} are multi-dimensional Gaussian integrals that we can evaluate straightforwardly
\begin{equation}
    \int d^{2(M-\vert Z\vert)}\bm{\beta} \, e^{-\bm{\beta}^\intercal[(\Sigma^{-1})_{(Z)}+D_{(Z)}]\bm{\beta}} = \frac{\pi^{M-\vert Z\vert }}{ \sqrt{\det{(\Sigma^{-1} )_{(Z)}+D_{(Z)}}}} \, .
    \label{gaussian_integral}
\end{equation}
Putting everything together we obtain
\begin{equation}
    p(\bm{k})  =\frac{1}{\sqrt{\det{\Sigma}}}  \sum_{0\leq \bm{d} \leq \bm{k}} 
    \prod_{i=1}^M \left[ \binom{N}{N-k_i,k_i-d_i,d_i}(-1)^{k_i-d_i} \right]\prod_{i\notin Z}\left[\frac{N}{d_i} \right]
    \frac{1}{\sqrt{\det{(\Sigma^{-1})_{(Z)}+D_{Z}}}}
    \label{probabilityv1} \, .
\end{equation}
To conclude the derivation we define the \textit{Kensingtonian}, a new matrix function whose action is specified once $N$ and $\bm{k}$ have been fixed. This allows us to express the probability as
\begin{equation}
    p(\bm{k}) = \frac{\text{Ken}[O]}{\sqrt{\det{\Sigma}}} \, ,
\end{equation}
where $O = \mathbb{I}-\Sigma^{-1}$ and 
\begin{equation}
    \text{Ken}[A] = \sum_{0\leq \bm{d} \leq \bm{k}}
     \prod_{i=1}^M \left[ \binom{N}{N-k_i,k_i-d_i,d_i}(-1)^{k_i-d_i} \right]\prod_{j\notin Z}\left[\frac{N}{d_j} \right]
     \frac{1}{\sqrt{\det{(\mathbb{I}-A)_{(Z)}+D_Z}}} \, .
\end{equation}

\section{Relation between the Kensingtonian and the Torontonian}
\label{appendix_ker_to_tor}
In this section we show that, starting from the expression for the click-counting probability distribution for a GBS experiment Eq.~\eqref{clickGBSprob}, we can retrieve the results of Ref.~\cite{Torontonian} by setting $N=1$.
In this scenario, there are only two possible measurement outcomes for each detector, namely $k_i = 0,1$. This in turn implies that the summation variable $d_i$ can either be equal to $0$ or $1$ and consequently that the matrix $D_{Z}$ vanishes, as can readily be seen from its definition Eq.~\eqref{Dmatrix}.
Eq.~\eqref{ken_def} then simplifies to
\begin{equation}
    \text{Ken}[A] = \sum_{0\leq \bm{d} \leq \bm{k}}
     \prod_{i=1}^M \left[(-1)^{k_i-d_i} \right]
     \frac{1}{\sqrt{\det{(\mathbb{I}-A)_{(Z)}}}} = \sum_{0\leq \bm{d} \leq \bm{k}}
     (-1)^{n-\sum_i d_i}
     \frac{1}{\sqrt{\det{(\mathbb{I}-A)_{(Z)}}}}
     \label{ken_to_tor_1}
\end{equation}
where $n=\sum_i k_i$ is the total number of clicks.
Let us also define two new sets for future convenience
\begin{equation}
    \mathcal{K}=\lbrace i\vert 1\leq i\leq M ,\, k_i=0  \rbrace \, ,
\end{equation}
\begin{equation}
    \mathcal{X}=\lbrace i\vert 1\leq i \leq M , \, d_i=0 \, \text{and} \, k_i=1 \rbrace \, .
\end{equation}
By noticing that $n-\sum_i d_i$ corresponds to the cardinality of $\mathcal{X}$ and that $Z = \mathcal{X}\cup\mathcal{K}$ we can rewrite Eq.~\eqref{ken_to_tor_1} as 
\begin{equation}
\begin{split}
    \text{Ken}[A]  =  \sum_{0\leq \bm{d} \leq \bm{k}}
     (-1)^{\vert \mathcal{X}\vert}
     \frac{1}{\sqrt{\det{(\mathbb{I}-A)_{(\mathcal{K}\cup \mathcal{X})}}}}
     = \sum_{Y\in P([n])}
     (-1)^{\vert \mathcal{Y} \vert}
     \frac{1}{\sqrt{\det{(\mathbb{I}-A_{(\mathcal{K})})_{(\mathcal{Y})}}}} \equiv \text{Tor}[A_{(\mathcal{K})}]\, ,
\end{split}
\end{equation}
where $P([n])$ is the power set of $[n]=\lbrace 1,\dots,n \rbrace$ and $\text{Tor}$ is the Torontonian of a matrix. 
Note how, strictly speaking, the Torontonian is defined slightly differently in Ref.~\cite{Torontonian}.
The reason lies in the fact that the authors used a different quadrature operators ordering and this choice is reflected in different rules for deleting rows and column from the matrix $A$.
Consequently, we should more accurately say that we have found an equivalent expression of the Torontonian valid for the operator ordering set by Eq.~\eqref{ordering}.
\\
We have thus shown that for $N=1$ we correctly retrieve the probability distribution obtained in Ref.~\cite{Torontonian}
\begin{equation}
    p(\bm{k}) = \frac{\text{Ken}[O]}{\sqrt{\det{\Sigma}}} = \frac{\text{Tor}[O_{(\mathcal{K})}]}{\sqrt{\det{\Sigma}}} \, .
\end{equation}
Lastly, note that the $\Sigma$ matrix used in Ref.~\cite{Torontonian} differs from the one used in this paper by a unitary transformation. This transformation, however, does not affect the determinants contained in the probability formula, leaving the latter unchanged.

\section{The loop-Kensingtonian}
\label{appendix_loop_ken}
In this section we generalize the formula for the click-counting detection probability distribution of an $M-$mode Gaussian state Eq.~\eqref{clickGBSprob}, when we lift the zero-displacement constraint.
This is achieved by introducing a new matrix function that we name the \textit{loop Kensingtonian}. 
The latter plays an analogous role to the loop Torontonian Ref.~\cite{looptorpaper} and the loop Hafnian Ref.~\cite{detailedGBS} functions in GBS setups employing threshold detection and PNR detection, respectively.
Apart from fundamental interest, this generalization is also motivated by the fact that some of the applications of GBS, like computing molecular vibronic spectra, require displacement in order to encode the problem into a Gaussian boson sampler.
\\
The derivation proceeds similarly to what we have seen in the zero-displacement case, the only difference being that the $Q$ function of the Gaussian state $\rho$ to be substituted in Eq.~\eqref{prob_Q_and_P_func} now reads
\begin{equation}
    Q_\rho (\bm{\beta}) =  \frac{1}{\pi^M\sqrt{\det{{\Sigma}}}}e^{-(\bm{\beta}-\bm{\alpha})^\intercal{\Sigma}^{-1}(\bm{\beta}-\bm{\alpha})} \, ,
\end{equation}
where $\bm{\alpha}$ is the displacement of $\rho$. Recall that our conventions are such that a single mode coherent state $\ket{\gamma}$ has displacement vector equal to $\bm{\alpha} = (\Re{\gamma},\Im{\gamma})^\intercal$. Following the same steps of Appendix \ref{appendix_ken_derivation} we get to 
\begin{equation}
\begin{split}
    p(\bm{k}) & =  \sum_{0\leq \bm{d} \leq \bm{k}} \prod_{i=1}^M \left[ \binom{N}{N-k_i,k_i-d_i,d_i} \frac{N(-1)^{k_i-d_i}}{\pi d_i}\right] \int d^{2M}\bm{\beta} \, \frac{ e^{-(\bm{\beta}-\bm{\alpha})^\intercal{\Sigma}^{-1}(\bm{\beta}-\bm{\alpha})}  }{\sqrt{\det{{\Sigma}}}} e^{ {\sum_{i=1}^M-\frac{N-d_i}{d_i}(\beta_{2i-1}^2+\beta_{2i}^2)}}  \\ & 
    = \sum_{0\leq \bm{d} \leq \bm{k}} \prod_{i=1}^M \left[ \binom{N}{N-k_i,k_i-d_i,d_i} \frac{N(-1)^{k_i-d_i}}{\pi d_i}\right]
    \frac{e^{-\bm{\alpha}^\intercal{\Sigma}^{-1}\bm{\alpha}}}{\sqrt{\det{\Sigma}}}
    \int d^{2M}\bm{\beta} \, e^{-\bm{\beta}^\intercal{\Sigma}^{-1}\bm{\beta}+2\bm{\alpha}^\intercal \Sigma^{-1}\bm{\beta}} e^{ {\sum_{i=1}^M-\frac{N-d_i}{d_i}(\beta_{2i-1}^2+\beta_{2i}^2)}} \, , 
 \end{split}
\end{equation}
where we have used the fact that $\Sigma^{-1}$ is a symmetric matrix. Integrating over the delta function contributions (i.e. $d_i = 0$) we obtain
\begin{equation}
    p(\bm{k})  =  \sum_{0\leq \bm{d} \leq \bm{k}}  \prod_{i=1}^M \left[ \binom{N}{N-k_i,k_i-d_i,d_i}(-1)^{k_i-d_i} \right]\prod_{i\notin Z}\left[\frac{N}{\pi d_i} \right]
    \frac{{e^{-\bm{\alpha}^\intercal{\Sigma}^{-1}\bm{\alpha}}}}{\sqrt{\det{\Sigma }}}
    \int d^{2(M-\vert Z\vert)}\bm{\beta}\,e^{-\bm{\beta}^\intercal[(\Sigma^{-1})_{(Z)}+D_{Z}]\bm{\beta}+\bm{\beta}^\intercal(2\Sigma^{-1}\bm{\alpha})_{(Z)}} \, ,
\end{equation}
where $Z$ and $D_Z$ are defined by Eq.~\eqref{Z_def} and Eq.~\eqref{Dmatrix}, respectively. Note that the subscript notation $(2\Sigma^{-1}\bm{\alpha})_{(Z)}$ applied to a vector denotes the elimination of its elements according to the set $Z$.
We can now easily evaluate the remaining Gaussian integrals in the previous expression 
\begin{equation}
    \int d^{2(M-\vert Z\vert)}\bm{\beta} \, e^{-\bm{\beta}^\intercal[(\Sigma^{-1})_{(Z)}+D_{Z}]\bm{\beta}+\bm{\beta}^\intercal (2\Sigma^{-1}\bm{\alpha})_{(Z)}}
    = \frac{\pi^{M-\vert Z\vert }}{\sqrt{\det{(\Sigma^{-1} )_{(Z)}+D_{Z}}}}
    e^{(\Sigma^{-1}\bm{\alpha})^\intercal_{(Z)}[(\Sigma^{-1})_{(Z)}+D_Z]^{-1}(\Sigma^{-1}\bm{\alpha})_{(Z)}}\, .
\end{equation}
Putting everything together we obtain
\begin{equation}
    p(\bm{k})  =\frac{e^{-\bm{\alpha}^\intercal\Sigma^{-1}\bm{\alpha}}}{\sqrt{\det{\Sigma}}}  \sum_{0\leq \bm{d} \leq \bm{k}} 
    \prod_{i=1}^M \left[ \binom{N}{N-k_i,k_i-d_i,d_i}(-1)^{k_i-d_i} \right]\prod_{j\notin Z}\left[\frac{N}{d_j} \right]
    \frac{e^{(\Sigma^{-1}\bm{\alpha})^\intercal_{(Z)}[(\Sigma^{-1})_{(Z)}+D_Z]^{-1}(\Sigma^{-1}\bm{\alpha})_{(Z)}}}{\sqrt{\det{(\Sigma^{-1})_{(Z)}+D_{Z}}}}\, .
\end{equation}
Finally, we can rewrite the previous expression as
\begin{equation}
    p(\bm{k}) = p(\bm{0})\text{lken}[O,\bm{\alpha}] \,  ,
\end{equation}
where $p(\bm{0})=e^{-\bm{\alpha}^\intercal\Sigma^{-1}\bm{\alpha}}/\sqrt{\det{\Sigma}}$, $O = \mathbb{I}-\Sigma^{-1}$ and 
\begin{equation}
    \text{lken}[A,\bm{\alpha}] = \sum_{0\leq \bm{d} \leq \bm{k}}
     \prod_{i=1}^M \left[ \binom{N}{N-k_i,k_i-d_i,d_i}(-1)^{k_i-d_i} \right]\prod_{j\notin Z}\left[\frac{N}{d_j} \right]
     \frac{{e^{((\mathbb{I}-A)\bm{\alpha})^\intercal_{(Z)}[(\mathbb{I}-A)_{(Z)}+D_Z]^{-1}((\mathbb{I}-A)\bm{\alpha})_{(Z)}}}}{\sqrt{\det{(\mathbb{I}-A)_{(Z)}+D_Z}}} \, .
\end{equation}
is the loop Kensingtonian of a matrix $2M\times 2M$ matrix $A$.

\section{Click-counting detection POVM in the $N\rightarrow\infty$ limit}
\label{Appendix_bigNlimit}
In this section we prove that in the $N\rightarrow\infty$ limit, ideal click-counting detection converges PNR detection. To do so, we simply Taylor-expand the POVM element Eq.~\eqref{click_POVM}
\begin{equation}
\begin{split}
    \Pi_k^{(N)} & = \,: \binom{N}{k} e^{-\hat{n}}   (e^{\frac{\hat{n}}{N}}-1)^k \, : = \,:\frac{N!}{k!(N-k)!} \left(\frac{\hat{n}}{N}\right)^k e^{-\hat{n}}: + \,O\left(\frac{1}{N}\right)  \\ & = 
     \frac{N(N-1)\cdots(N-k+1)}{N^k} :\frac{\hat{n}^k}{k!} e^{-\hat{n}}: + \,O\left(\frac{1}{N}\right) \stackrel{\small{N\rightarrow\infty}}{\rightarrow }\,:\frac{\hat{n}^k}{k!} e^{-\hat{n}}: \,\equiv \ketbra{k} \, .
\end{split}
\end{equation}

\section{Complexity of click-GBS (alternative proof)}
\label{Appendix_alternative_proof}
In what follows we present an alternative proof of hardness for click-counting GBS that does not rely on using the output probability distribution of a GBS task employing threshold detectors.
For simplicity of exposition, we consider a single mode scenario and let $\rho$ be the output state of the LON. 
We show that, in the non-collisional regime, the output probability distribution $p$ of click-counting GBS is arbitrarily close (in total variational distance) to that of a Gaussian boson sampler employing PNR detectors, which we denote with $\tilde{p}$.
The latter reads $\tilde{p}(k)=\Tr\lbrace\rho\tilde{\Pi}_k\rbrace$ for $k\in\lbrace 0,1,\dots\rbrace$, and the POVM element is a projector on a Fock state, i.e. $\tilde{\Pi}_{k}=\ketbra{k}$.
For click-counting detection we have $p(k)=\Tr\lbrace\rho\Pi^{(N)}_k\rbrace$ with $k \in\lbrace 0,\dots,N\rbrace$, and the POVM element $\Pi_k^{(N)}$ is given by Eq.~\eqref{click_POVM}.
The latter may be alternatively expressed as \cite{clickdetection}
\begin{equation}
    \Pi^{(N)}_k = \binom{N}{k} \sum_{n=k}^\infty \frac{1}{N^n}\partial_x^n[e^x-1]^k\vert_{x=0}\ketbra{n} \equiv \sum_{n=k}^\infty c_k(n) \ketbra{n} = c_k(k) \ketbra{k} + \sum_{n=k+1}^\infty c_k(n) \ketbra{n} \, ,
\end{equation}
where
\begin{equation}
    c_k(n) = \binom{N}{k}\frac{1}{N^n}\partial_x^n[e^x-1]^k\vert_{x=0} \, .
\end{equation}
Notice how for $k>N$ we can set $\Pi_k^{(N)}=0$ without loss of generality. 
The total variational distance between $p$ and $\tilde{p}$ reads 
\begin{equation}
    \vert\vert \tilde{p}-p \vert\vert_1 = \frac{1}{2}\sum_{k=0}^\infty \vert  \tilde{p}(k)-p(k) \vert  = \frac{1}{2}\sum_{k=0}^\infty \vert  \Tr\lbrace\rho(\tilde{\Pi}_k-\Pi_{k}^{(N)})\rbrace \vert  \, .
\end{equation}
We can then write $\tilde{\Pi}_k-\Pi_k^{(N)} = A_k - B_k$, where
\begin{equation}
    A_k = \begin{cases}
     (1-c_k(k))\ketbra{k} & 0\leq k\leq N \\
     \ketbra{k} & k>N
    \end{cases}
    \, , \quad\quad B_k = \begin{cases}
     \sum_{n=k+1}^\infty c_k(n) \ketbra{n} & 0\leq k\leq N \\
     0 & k>N
    \end{cases} \, , 
\end{equation}
and 
\begin{equation}
    c_k(k) = \frac{N!}{(N-k)!N^k} \, .
    \label{coefficients}
\end{equation}
Since the POVM elements sum to the identity $\sum_{k=0}^\infty \tilde{\Pi}_k = \sum_{k=0}^\infty {\Pi}^{(N)}_k = \mathcal{I}$, we have that
\begin{equation}
    \sum_{k=0}^\infty (\tilde{\Pi}_k-\Pi_k^{(N)}) = \sum_{k=0}^\infty (A_k - B_k) = 0 \, ,
\end{equation}
which in turn implies that $\sum_{k=0}^\infty A_k = \sum_{k=0}^\infty B_k$.
Using this and the fact that both $A_k$ and $B_k$ are positive semi-definite operators, we can write 
\begin{equation}
\begin{split}
    \vert\vert \tilde{p}-p \vert\vert_1 & = \frac{1}{2}\sum_{k=0}^\infty \vert \Tr{\rho A_k} -\Tr{\rho B_k}\vert \leq
    \frac{1}{2}\sum_{k=0}^\infty \left( \Tr{\rho A_k} + \Tr{\rho B_k} \right) = \sum_{k=0}^\infty \Tr{\rho A_k} \\ &
    = \sum_{k=0}^{N} (1-c_k(k)) \bra{k}\rho\ket{k} +\sum_{k=N+1}^{\infty} \bra{k}\rho\ket{k}  = 
    \sum_{k=0}^{N}(1-c_k(k))\tilde{p}(k)+\sum_{k=N+1}^{\infty}\tilde{p}(k) \, .
\end{split}
\end{equation}
Notice how the $k=0$ and $k=1$ terms are actually null, since Eq.~\eqref{coefficients} implies that $c_0(0) = c_1(1) = 1$.
In the non-collisional regime, the probability $\varepsilon$ of observing two or more photons
\begin{equation}
    \varepsilon = \sum_{k=2}^\infty \tilde{p}(k) 
\end{equation}
becomes negligible.
Hence, we can finally write 
\begin{equation}
    \vert\vert \tilde{p}-p \vert\vert_1 \leq \varepsilon \, ,
\end{equation}
ans use the same arguments presented in the main text to claim the computational complexity of click-counting GBS.
The generalization to the $m-$mode, although cumbersome, may be easily obtained following the steps outlined above.

\end{document}